\documentclass[a4paper]{article}
\def\email#1{\thanks{\texttt{#1}}}
\def\institute#1{\date{#1}}
\def\keywords#1{}
\usepackage[english]{babel}
\usepackage[latin1]{inputenc}
\usepackage[T1]{fontenc}
\usepackage{amssymb}
\usepackage{eucal}
\usepackage{booktabs}
\usepackage[matrix,arrow,line]{xy} \CompileMatrices

\let\phi=\varphi
\let\epsilon=\varepsilon
\let\e=\epsilon

\def\Z{\mathbb Z}
\def\P{\mathcal{P}}
\def\S{\mathcal{S}}

\def\dom{\mathop{\rm dom}}

\def\op{{\oplus}}
\def\om{{\ominus}}

\def\msmaller{\mathchoice{\scriptstyle}{\scriptstyle}{\scriptscriptstyle}{}}
\def\0{{\msmaller\square}}
\def\1{{\msmaller\blacksquare}}

\def\set#1{\{\,#1\,\}}

\def\ifdisplay#1#2{\mathchoice{#1}{#2}{#2}{#2}}

\let\longto=\longrightarrow
\let\shortto=\to
\def\to{\ifdisplay{\longto}{\shortto}}

\let\shortmapsto=\mapsto
\def\mapsto{\ifdisplay{\longmapsto}{\shortmapsto}}

\newcommand\rea[1][{}]{\rightarrow_{#1}}

\let\text=\mbox
\def\qtext#1{\quad\text{#1}\quad}    
\def\qqtext#1{\qquad\text{#1}\qquad} 

\def\smath{\everymath={\scriptstyle}}
\def\basesituation#1{%
  {\let\\=\cr \baselineskip=1.2ex \lineskip=0.4ex \lineskiplimit=1pt
    \halign{\hfil$##{}$&&\hfil${}##$\hfil\cr #1 \crcr}}}

\def\situation#1{{\smath\vcenter{\basesituation{#1}}}}
\def\bsituation#1{\vbox{\basesituation{#1}}} 

\def\reaction#1#2{\vcenter{\basesituation{&#2&\\#1}}}
\def\sreaction#1#2{{\smath \reaction{#1}{#2}}}

\def\eqref#1{(\ref{#1})}

\begin{document}
\title{Flexible Time and the Evolution of One-Dimensional Cellular
  Automata}
\author{Markus Redeker\email{cep@ibp.de}}
\institute{Hamburg, Germany}
\maketitle

\begin{abstract}
  Here I describe a view of the evolution of cellular automata that
  allows to operate on larger structures.  Instead of calculating the
  next state of all cells in one step, the method here developed uses
  a time slice that can proceed at different places differently.  This
  allows to ``jump'' over the evolution of known structures in a
  single step.
\end{abstract}

\section{Introduction}

In this text I introduce a generalised time concept which is helpful
for the study of cellular automata. It is motivated by the view of a
cellular automaton as a parallel computer which executes a number of
tasks that require different amounts of time. The relevant time
concept for the description of such computations is therefore not
clock time or the number of executed machine instructions but the
moment when a certain part of the computation has been
completed.\footnote{The earliest form of a similar idea of which I
  know occurs in~\cite{Gosper1984}. Or you may view it as an
  application of the concept of ``many-fingered time'' in General
  Relativity \cite[p. 714]{MisnerThorneWheeler1973} to cellular
  automata.}

This ``moment'' is a snapshot of the computation and contains
different parts of the machine at different times. For one-dimensional
cellular automata we get a snake-line picture like, e.\,g.
$\situation{
      &\0&\\
      &\0&\0&  &  &        &  &    &\0&\\
      &\0&  &\0&  &        &\0&\1\1&\0&\0\1\\
    \1&\0&  &  &\0&        &\0&\\
      &  &  &  &  &\0\1\1\1&\0&}$\,.

In this text I describe a system that allows one to work with such
objects. I will introduce three related variants; the last one is an
example of how the behaviour of several cells over a number of time
steps can be described as a single operation: In the computation
analogy, we have identified a very simple subroutine.

\section{Cellular Processes}

The first concept I introduce is that of a \emph{cellular process},
which describes the behaviour of some cells at some time.

Let $\Sigma$ be the set of the \emph{states} that a cell may have.
Then a cellular process with states in $\Sigma$ is simply a function
\begin{equation}
  \label{eq:25}
  \pi\colon W \to \Sigma
\end{equation}
with $W \subseteq \Z^2$. The set of all such processes is
$\P(\Sigma)$.

$W$ is the usually finite observation window to the behaviour of an
infinite line of cells. Its elements are space-time points of the form
$(t,x)$, where $x$ is the position of a cell and $t$ the time at which
it is observed. Its state at this time is then $\pi(t,x)$.

A cellular process is also a formalisation of diagrams like
$\situation{
  \1\1\1\1\1\\
  \0\1\1\0\1\\
  \0\0\1\1\1\\
  \0\0\0\1\1}$\,. This one has $\Sigma = \{\0, \1\}$, and $W$ may be
$\{0, 1, 2, 3\} \times \{0, 1, 2, 3, 4\}$. I use in this text the
convention that time runs \emph{upwards}, therefore the line at the
bottom contains the earliest generation of cells and the whole diagram
shows their behaviour over four time steps. Equations for $\pi$ can
then be read directly from the diagram, $\pi(0,0) = \pi(1,0) =
\pi(2,0) = \0$, $\pi(3,0) = \1$, and more. The four values of $\pi$
that I just wrote down describe the states of the cell at $x = 0$ over
four time steps.

The reason for definition~\eqref{eq:25} is that it gives us
automatically an arithmetic for cellular processes. We must only view
processes like $\pi$ set-theoretically, as a set of pairs $(p,
\pi(p))$, then expressions like $\pi \subseteq \theta$, $\pi \cap
\theta$, and $\pi \cup \theta$ have obvious meanings for all cellular
processes $\pi$ and $\theta$. This also means that $\emptyset$ is the
process with empty domain and that the set $W$ in
equation~\eqref{eq:25} needs no name of its own but can be written as
$\dom \pi$.

However, the union $\pi \cup \theta$ of two cellular processes is not
always a function: there might be a $p \in \dom\pi \cap \dom\theta$
with $\pi(p) \neq \theta(p)$. If there is no such $p$, then $\pi$ and
$\theta$ are \emph{compatible}. Operations on cellular processes will
usually be restricted to compatible ones.

\section{Transition rules}

A special kind of cellular processes describe the behaviour of
cellular automata. In them, the state of a cell at time $t+1$ depends
on its own state and that of a finite number of neighbours at time
$t$.

If the state of a cell is determined by its $r$ nearest neighbours at
each side, we have a \emph{transition function}
\begin{equation}
  \phi\colon \Sigma^{2r+1} \to \Sigma,
\end{equation}
and $r$ is the \emph{radius} of the automaton.

We have then for every element of $\Sigma^{2r+1}$ a cellular process
\begin{equation}
  \label{eq:16}
  \def\arraystretch{0.7}
  \begin{array}[t]{rll}
                      & \tau \\
    \sigma_{-r} \dots & \sigma_0 & \dots \sigma_r
  \end{array}
\end{equation}
which expresses that a cell in state $\sigma_0$ is in the next time
step in state $\tau$ if its left neighbours are in states
$\sigma_{-r}$, \dots, $\sigma_{-1}$ and its right neighbours are in
states $\sigma_1$, \dots, $\sigma_r$. With $\phi$ this would be
written as $\phi(\sigma_{-r} \dots \sigma_r) = \tau$.

A list of processes in the form~\eqref{eq:16} gives therefore a very
visual way to write $\phi$. But it tells almost nothing about the
global behaviour of the cellular automaton; we will have to rewrite it
to get something understandable.

\subsection{Rule 110 and a computation}

The concrete question that motivates this all is about the behaviour
of the elementary cellular automaton 110 in Stephen Wolfram's
numbering scheme~\cite{Wolfram1983}, commonly called \emph{Rule~110}.
It has $r=1$, $\Sigma = \{0,1\}$, and
\begin{equation}
  \phi(w) =
  \left\{
      \begin{array}{r@{\quad}l}
        0&\text{for $w \in \{000, 100, 111\}$,}\\
        1&\text{for all other $w \in \Sigma^3$.}
      \end{array}
    \right.
\end{equation}

The diagrams for them in the style of~\eqref{eq:16} look much clearer
if we write them with $\0$ and $\1$ instead of 0 and 1:
\begin{equation}\label{eq:2}
  \sreaction{\0&\0&\0}{\0}, \quad
  \sreaction{\0&\0&\1}{\1}, \quad
  \sreaction{\0&\1&\0}{\1}, \quad
  \sreaction{\1&\0&\0}{\0}, \quad
  \sreaction{\0&\1&\1}{\1}, \quad
  \sreaction{\1&\0&\1}{\1}, \quad
  \sreaction{\1&\1&\0}{\1}, \quad
  \sreaction{\1&\1&\1}{\0}\,.
\end{equation}
With them we can compute the evolution of a cell configuration
directly. We may start, e.\,g., with the line
$\situation{\1\0\0\0\1\0\0\1}$ and extend it to
$\sreaction{\1&\0\0\0\1\0\0&\1}{\0\0\1\1\0\1}$ by placing over every
subsequence of three cells the top cell in the corresponding diagram
in~\eqref{eq:2}. The new cells form another line,
$\situation{\0\0\1\1\0\1}$, which can be extended in the same way. We
stop here and get a computation in three steps,
\begin{equation}
  \label{eq:26}
  \1\0\0\0\1\0\0\1
  \rea \bsituation{  \0&\0\1\1\0\1&\\
                   \1\0&\0\0\1\0\0&\1}
  \rea \bsituation{    &\1\1\1\1&\\
                     \0&\0\1\1\0&\1\\
                   \1\0&\0\0\1\0&\0&\1}\,.
\end{equation}
Each step in~\eqref{eq:26} is itself a partial computation of the
cellular automaton and extends the previous one. Now remove the
repeated parts so that only the end situations are left,
\begin{equation}\label{eq:3}
  \1\0\0\0\1\0\0\1
  \rea \bsituation{\0&\0\1\1\0&\1&\\
                   \1\0&&\0&\1}
  \rea \bsituation{&\1&\1\1&\1&&\\\0&\0&&\0&\1\\\1\0&&&&\0&\1}\,.
\end{equation}
The resulting cellular processes are almost linear sequences of cells,
only a little bit bent. With an appropriate notation they can even be
written as lines of cells, namely as
\begin{equation}\label{eq:6}
  10001001
  \rea 10 \op 00 11 01 \om 01
  \rea 10 \op 00 \op 1111 \om 01 \om 01,
\end{equation}
where I have written the states of the cells once again as numbers to
let it look even more like a formula.

This is the system for the description of cellular automata which I
will now develop in detail. We will call~\eqref{eq:6} a sequence of
two \emph{reactions} (written as arrows) between three
\emph{situations}. These terms are explained in the next section. To
distinguish between situations and cellular processes, I will use the
symbols $\0$ and $\1$ only for processes and the digits 0 and 1 only
for situations.

\section{Situations}

A \emph{situation} is a sequence of cell states and certain elements
$[p]$, which represent gaps between the cell states.

I will now describe these two kinds of situations and then define a
product between situations: the set of all situations consists of all
finite products of the elementary ones.

\subsection{Elementary Situations}
\label{sec:elementary-processes}

To every situation $a$ belongs a \emph{size}, $\delta(a) \in \Z^2$,
and a cellular process $\pi_a$. Conceptually, $a$ is a sequence of
cells that reaches from $(0,0)$ at the left to $\delta(a)$ at the
right (even then if $\pi_a$ has no cell at $(0,0)$ or $\delta(a)$).
$\delta(a)$ is the difference between the end and the start of the
cell sequence $a$, therefore the symbol $\delta$.

We start with the two kinds of \emph{elementary situations}.
\begin{enumerate}
\item Every \emph{cell state} $\sigma \in \Sigma$ is a situation, with
  \begin{equation}
    \begin{array}[t]{rl}
      \pi_\sigma\colon \{ (0,0) \} &\longto \Sigma\\
      p\  & \longmapsto \sigma
    \end{array}
    \qqtext{and} \delta(\sigma) = (0,1)\,.
  \end{equation}
  It describes the case that the cell at $x=0$ at time $t=0$ is in
  state $\sigma$.

\item There is for every $p \in \Z^2$ a different situation $[p]$, a
  \emph{displacement}. The set of displacements is also disjoint from
  $\Sigma$, and we have
  \begin{equation}
    \pi_{[p]} \colon \emptyset \longto \Sigma
    \qqtext{and} \delta([p]) = p\,.
  \end{equation}
  Since $\pi_{[p]}$ is empty, a displacement tells nothing about the
  cells, but it is useful to manipulate $\delta$ values. We may
  abbreviate $[(t,x)]$ as $[t,x]$ and $[0, x]$ as $[x]$.
\end{enumerate}
Displacements of this general form appear seldom in this text. We will
mostly use the definitions
\begin{equation}
  \om_i = [-1, -i] \qqtext{and} \op_i = [1, -i]\,.
\end{equation}
With a cellular automaton of radius $r$ we will abbreviate further and
use the forms $\om = \om_r$ and $\op = \op_r$.

\subsection{Products}

All situations are \emph{products} of the elementary ones. The product
is subject to a compatibility condition.
\begin{enumerate} \setcounter{enumi}{2}
\item If $[p]$ is a displacement and $a$ an arbitrary situation, their
  product $[p]a$ has
  \begin{equation}
    \begin{array}[t]{rcl}
      \pi_{[p]a} \colon &p + \dom \pi_a &\longto \Sigma\\
      &p + q  & \longmapsto \pi_a(q)
    \end{array}
    \text{and}\qquad
    \delta([p]a) = p + \delta(a)
  \end{equation}
  and is a copy of $a$ that is shifted by $p$.

\item The product $ab$ of two situations $a$ and $b$ is then defined
  by
  \begin{equation}
    \pi_{ab} = \pi_a \cup \pi_{[\delta(a)]b}
    \qqtext{and} 
    \delta(ab) = \delta(a) + \delta(b)
  \end{equation}
  and exists if $\pi_a$ and $\pi_{[\delta(a)]b}$ are compatible. It
  consists of a shifted version of~$b$ attached to the right end of
  $a$.

\item The set of all finite products of the elementary situations,
  with the empty product written as $\lambda$, is $\S(\Sigma)$.
\end{enumerate}

Two situations $a$ and $b$ are \emph{compatible} if $\pi_a$ and
$\pi_b$ are compatible and $\delta(a) = \delta(b)$. We can therefore
say that $ab$ exists if $a[\delta(b)]$ and $[\delta(a)]b$ are
compatible.

\subsection{Sets of Situations}

Now that we have products, all conventions for then can be used.
Therefore $a^k$ is the $k$-th power of $a$, and $a^0 = \lambda$. The
set of all powers of $a$, with or without~$\lambda$, is $a^* =
\set{a^k: k \geq 0}$ or $a^+ = \set{a^k: k > 0}$, respectively. For a
set $S$ of situations exist the multiplicative closures $S^*$ and
$S^+$, where $S^+$ is the set of all products of elements of $S$, and
$S^* = S^+ \cup \{\lambda\}$. Note that all the products involved are
subject to a compatibility condition; it is therefore, e.\,g.,
possible that $a^*$ is finite.

Because the elementary situations in an $a \in \S(\Sigma)$ have a
fixed order, it is meaningful to speak of the factors of $a$. If there
are situations $a_1$ and $a_2$ with $a = a_1 b a_2$, then $b$ is a
\emph{factor} of $a$. This will be used to define situations in terms
of forbidden factors.

A third concept that has turned out to be very useful for the
definition of situations are \emph{extension rules}. Let $a$ and $m$
be situations and $M$ a set of situations. Then we say that \emph{in
  $a$, $m$ extends to $M$} if for every decomposition
\begin{equation}
  a = a_1 m a_2
\end{equation}
we have a decomposition
\begin{equation}
  a = b_1 m_1 m m_2 b_2
\end{equation}
with $a_1 = b_1 m_1$, $a_2 =m_2 b_2$, and $m_1 m m_2 \in M$.

In this text I use the convention that $\rho$, $\sigma$, $\tau$,
\dots\ are elements of $\Sigma$, while $u$, $v$, $w$,~\dots\ are
elements of $\Sigma^*$, and $a$, $b$, $c$,~\dots\ are elements of
$\S(\Sigma)$.

\section{Reactions}

A \emph{reaction} is simply a pair $(a, a')$ of compatible situations.
It expresses the fact that in a certain cellular automaton the
situation $a$ is a result of the initial condition $a'$. In the
extended time concept of the introduction, $a'$ is ``later'' than~$a$.

The behaviour of a cellular automaton is then described by a
\emph{reaction system} $(S, \rea)$ that consists of a set $S$ of
situations and the reactions between them. The reactions, a subset of
$S \times S$, form a binary relation $\rea$, and if there is a
reaction $(a, a')$ that belongs to the system $(S, \rea)$, it is
written as $a \rea a'$.

A reaction system $(S, \rea)$ must obey the following rules:
\begin{enumerate}
\item if $a \in S$ then $a \rea a$, \emph{(Reflectivity)}
\item if $a \rea b$ and $b \rea c$ then $a \rea c$,
  \emph{(Transitivity)} and
\item if $b \rea b'$ and $abc \in S$, then $abc \rea ab'c$ and $ab'c
  \in S$. \emph{(Extension Rule)}
\end{enumerate}
As with other mathematical structures, I will write $(S, \rea)$ as $S$
if the context is unambiguous.

The first two conditions make $\rea$ a quasiorder on $S$. The third
one allows to define a reaction system by a small set of
\emph{generator reactions} and some initial situations. In the
simplest cases, the generator reactions are derived directly from
$\phi$ and describe the computation of exactly one new cell state.

If a reaction $abc \rea ab'c$ has been derived with rule 3, we say
that $b \rea b'$ has been \emph{applied} to $abc$. Note however that a
product like $ab'c$ needs not to exist and that therefore the
extension rule places an implicit condition on the reaction system and
it must be proved to be consistent.

\subsection{A General Reaction System}

The rewriting of $\phi$ to get something better understandable can now
begin with the construction of the reaction system $\Phi$. Its
generator reactions are
\begin{equation}
  \label{eq:wide-generators}
  \begin{array}[b]{rcl@{\qquad\qquad}rcl}
    \om \sigma w &\rea& \phi(\sigma w) \om w, &
    w \sigma \op &\rea& w \op \phi(w \sigma),\\
        w        &\rea& w {\op\om} w, &
    \om w \op    &\rea& \lambda\,.
  \end{array}
\end{equation}
for every $w \in \Sigma^{2r}$ and $\sigma \in \Sigma$. The set of all
situations contains all $b \in (\Sigma \cup \{\om, \op\})^*$ such that
\begin{enumerate}
\item in $b$, $\om$ extends to $\om \Sigma^{2r}$ and $\op$ extends to
  $\Sigma^{2r} \op$, and
\item there are $a$, $c \in \S(\Sigma)$ and $w \in \Sigma^*$ with $w
  \rea abc$.
\end{enumerate}
One can see that the generator reactions (and therefore all reactions)
preserve these conditions. The second condition is not absolutely
necessary for a consistent reaction system, but it makes the
consistency proof more easily generalisable.

\subsection{How it is Used}
\label{sec:how-used}

As an example how this system works, I will now show how the first
reaction in~\eqref{eq:6} is derived. The initial situation,
$10001001$, contains neither $\om$ nor $\op$, therefore we must first
apply a reaction of the type $w \rea w \op\om w$ to it, e.\,g. with $w
= 10$. We get then
\begin{equation}
  \label{eq:4}
  10001001 \rea 10 \op\om 10001001
\end{equation}
among other possibilities. One reaction from the top left
of~\eqref{eq:wide-generators}, namely $\om 100 \rea 0 \om 00$, can
then be applied to it, resulting in
\begin{equation}
  10 \op\om 10001001 \rea 10 \op 0 \om 0001001,
\end{equation}
and then others until we reach $10 \op 001101 \om 01$.

The last reaction type in~\eqref{eq:wide-generators}, $\om w \op \rea
\lambda$, is needed after reactions have been started from different
places. As an example, we could have continued after~\eqref{eq:4} with
a reaction to $10 \op\om 10001001 \op\om 01$, and later reached $10
\op 001101 \om 01 \op\om 01$ in the same way as before. The reaction
$\om 01 \op \rea \lambda$ then removes the extra factor $\om 01 \op$
and we get $10 \op 001101 \om 01$ again.

\subsection{The Unique Future}
\label{sec:unique-future}

We still have to answer the question whether the extension rule holds
for $\Phi$.

To do this, I define now for each $b \in \Phi$ a cellular process
$\bar\pi_b$, the \emph{future} of $b$. It contains $\pi_b$ and all the
cell states that are influenced by it through $\phi$. More precisely,
$\bar\pi_b$ is the smallest process (in the set-theoretic sense) that
has $\pi_b$ as a subset and where for every $p \in \Z^2$ and $w \in
\Sigma^{2r+1}$ we have
\begin{equation}
  \label{eq:wide-future}
  \text{if}\quad
  \pi_{[p]\om w} \subseteq \bar\pi_b
  \qtext{then}
  \pi_{[p]\phi(w)} \subseteq \bar\pi_b\,.
\end{equation}
Each cell state in $\bar\pi_b \setminus \pi_b$ depends therefore
uniquely on $2r + 1$ cells in the previous step; when $\bar\pi_b$
exists, it is by induction unique.

The following diagram is an example for $r=1$, with $\dom \pi_b$ shown
as $\circ$ and $\dom \bar\pi_b$ as $\cdot$ and~$\circ$\,:
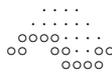
\begin{displaymath}
  \def\o{\circ&} \def\.{\cdot&}
  \situation{
    & & & & \.\.\\
    & & & \.\.\.\.\\
    & & \.\.\.\.\.\.\\
    & \o\o\o\o\.\.\.\.\\
    \o\o & & \o\o\.\.\o\o\\
    & & & & & \o\o\o\o\\
  }
\end{displaymath}

The future of a situation is defined in this way because we have then
\begin{equation}
  \label{eq:wide-rea-future}
  \text{if}\quad
  b \rea b'\
  \qtext{then}
  \bar\pi_b \supseteq \bar\pi_{b'}
\end{equation}
provided that $\bar\pi_b$ exists; in this case $\bar\pi_{b'}$ exists,
too. The proof of this begins with the generator reaction
in~\eqref{eq:wide-generators}, which
fulfil~\eqref{eq:wide-rea-future}. Now if $b \rea b'$ is a generator
reaction, $abc$ is an element of $\Phi$, and $\bar\pi_{abc}$ exists,
then $\bar\pi_{abc} \supseteq \bar\pi_{[\delta(a)]b}$ and
$\bar\pi_{[\delta(a)]b} \supseteq \bar\pi_{[\delta(a)]b'}$
(by~\eqref{eq:wide-rea-future}), therefore $\bar\pi_{abc} \supseteq
\bar\pi_{ab'c}$. This means that~\eqref{eq:wide-rea-future} is true
for reactions where a single generator reaction is applied to a
situation. By induction it is therefore true for all reactions in
$\Phi$.

With~\eqref{eq:wide-rea-future} we can now see that in fact every $b
\in \Phi$ has a future. This is because every $w \in \Sigma^*$ has a
future and there is for every $b \in \Phi$ a reaction $w \rea abc$. So
we have $\bar\pi_w \supseteq \bar\pi_{abc} \supseteq
\bar\pi_{[\delta(a)]b}$, therefore $\bar\pi_{[\delta(a)]b}$ has a
future. But this is only a shifted version of $\bar\pi_b$. (Instead of
$w \in \Sigma^*$ I could have used in the definition of $\Phi$
elements of a larger set for which a future exists, but for the
present purpose $\Sigma^*$ is enough.)

The extension rule, finally, is a side effect of the proof
of~\eqref{eq:wide-rea-future}. We have already seen that it is valid
for generator reactions: If $b \rea b'$ is a generator reaction and
$abc \in \Phi$, then $\bar\pi_{abc} \supseteq \bar\pi_{ab'c}$ and
therefore $\bar\pi_{ab'c}$ exists. But then it exists by induction for
every reaction $b \rea b'$.

Other facts that follow from~\eqref{eq:wide-rea-future} are: If $a
\rea b$ and $a \rea b'$ then $b$ and $b'$ are compatible, and if also
$\dom \pi_b = \dom \pi_{b'}$, then $\pi_b = \pi_{b'}$. This means that
different reaction paths, as in Section~\ref{sec:how-used}, lead to
essentially the same result.

\section{Narrow Rules}

The definition of $\bar\pi_b$ imitates the computation of a cell state
in a cellular automaton from the $2r +1$ states in its neighbourhood
one time step earlier.

But often not all of them are actually needed. In Rule 110, e.\,g., we
have both $\phi(000) = 0$ and $\phi(100) = 0$, so we need to know only
the two cells at the left to compute the next cell state. In other
words, the situation $00$, which has under
definition~\eqref{eq:wide-future} only the trivial future $\small
\0\0$, ``should have'' the future~$\sreaction{&\0&\0}\0$.

\subsection{A Better Future}

For this we need a more complex definition
than~\eqref{eq:wide-future}. In the new kind of future, $\hat\pi_b$,
the variable $w$ of~\eqref{eq:wide-future}, which represents the
predecessors of the cell state at $p$, is replaced with a whole set
\begin{equation}
  W_p = \set{ w \in \Sigma^{2n+1}:
    \pi_{[p]\om w} \text{ is compatible to } \hat\pi_b}
\end{equation}
of possible predecessor sequences. With them I define $\hat\pi_b$ as
the smallest process containing $\pi_b$ where for every $p \in \Z^2$
we have
\begin{equation}
  \label{eq:narrow-future}
  \text{if}\quad
  \exists \sigma \in \Sigma\,
  \forall w \in W_p\colon\, \phi(w) = \sigma
  \qtext{then}
  \pi_{[p]\sigma} \subseteq \hat\pi_b\,.
\end{equation}
Then every cell state in $\hat\pi_b \setminus \pi_b$ is uniquely
determined by the known part of its $2r + 1$ predecessors. Therefore,
as in the case of $\bar\pi_b$, if $\hat\pi_b$ exists, it is uniquely
determined by $b$.

Definition~\eqref{eq:narrow-future} can behave differently
from~\eqref{eq:wide-future} only on the boundaries of $\hat\pi_b$; we
get therefore longer and narrower diagrams than with~$\bar\pi$. For
example, $\hat\pi_{1^6}$ under Rule 110 is now
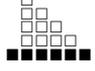
\begin{displaymath}
  \situation{
      &\0\\
      &\0&\0\\
      &\0&\0&\0\\
      &\0&\0&\0&\0\\
    \1&\1&\1&\1&\1&\1\,.\\
  }
\end{displaymath}
The corresponding reaction system $\Psi$, which I describe next, is
therefore called the \emph{narrow form} of $\Phi$.

\subsection{Defining Reactions}

The reaction system $\Psi$ is a generalisation of $\Phi$. Its
definition is more complex than that of $\Phi$, but its properties are
already given by a subset of its generator reactions. This subset, the
\emph{defining reactions} of $\Psi$, contains reactions of the form
\begin{equation}
  \label{eq:def-reactions}
  \om_i u \rea v \om_j w
  \qqtext{or}
  u' \op_{i'} \rea w' \op_{j'} v',
\end{equation}
with $u$, $u'$, \dots, $w' \in \Sigma^*$ and $i$, $i'$, $j$, $j' \in
\Z$.

Before $\Psi$ can be described, two sets of \emph{boundary terms},
$B_-$ and $B_+$, must be introduced. To define $B_-$, we use
temporarily a set $\tilde B_-$ that contains all terms $\om_j w$ for
which there is a defining reaction of the form $\om_i u \rea v \om_j
w$. Then $B_-$ is the set of ``shortest'' elements of $\tilde B_-$:
\begin{equation}
  B_- = \tilde B_- \setminus
  \set{ \om_j wx :  \om_j w \in \tilde B_-, x \in \Sigma^+ }\,.
\end{equation}
The set $B_+$ is constructed in the same way from the terms $w'
\op_{j'}$ that occur in the defining reactions at the right side
of~\eqref{eq:def-reactions}.

\medskip
Now we can define $\Psi$. Its situations are all $b \in (\Sigma \cup
\set{\om_i, \op_i : i \in \Z})^*$ where
\begin{enumerate}
\item in $b$, all $\om_i$ extend to $B_-$ and all $\op_i$ extend to
  $B_+$, and
\item there are $a$, $c \in \S(\Sigma)$ and $w \in \Sigma^*$ with $w
  \rea abc$.
\end{enumerate}
The generator reactions of $\Psi$ are then those
in~\eqref{eq:def-reactions} together with all minimal reactions of
the form
\begin{equation}
  \label{eq:narrow-open}
  w \rea u \op_i \om_j v
\end{equation}
with $u$, $v$, $w \in \Sigma^*$ and certain reactions of the form
\begin{equation}
  \label{eq:narrow-close}
  \om_i u \op_j \rea v
\end{equation}
with $u$, $v$, $w \in \Sigma^*$.

A reaction~\eqref{eq:narrow-open} is minimal if no reaction of the
same type can be applied to a part of $w$. The reactions
of~\eqref{eq:narrow-close} are constructed from all pairs of
reactions, $w \rea x \op_k\om_i u \op_j\om_{k'} x'$ and $w \rea x
\op_k v \om_{k'} x'$ with $w$, $x$, $x' \in \Sigma^*$, where either no
reaction of~\eqref{eq:def-reactions} can be applied to $\om_i u
\op_j$ or it would lead to a result with overlappings (like $v'
\om_{i'} u' \op_j$ with $u' \in \Sigma^\ell$ and $\ell < i' + j$).

\subsubsection{Subsystems} Two subsystems of $\Psi$ are sometimes
useful, $\Psi_-$ and $\Psi_+$. $\Psi_-$ is the system that has only
the reactions at the left side of~\eqref{eq:def-reactions} as defining
reactions. Therefore no reactions of the types described in
\eqref{eq:narrow-open} and~\eqref{eq:narrow-close} are possible for
it. The situations in $\Psi_-$ are those elements of $\Psi$ that
contain no $\op_i$:
\begin{displaymath}
  \Psi_- = \Psi \cap (\Sigma \cup \set{\om_i\colon i \in \Z})^*\,.
\end{displaymath}
Similarity, $\Psi_+$ is the reaction system that we get when we start
with the reactions at the right side of~\eqref{eq:def-reactions}.

\subsection{Narrowing}

Now we construct the defining reactions of $\Psi$. The construction
uses a sequence of intermediate reaction systems $\Psi_i = (\Psi_i,
\rea[i])$ with $\Psi_0 = \Phi$ and $\Psi_4 = \Psi$. It can be
described by transformations of the defining reactions because
everything else follows from them. I will only show the
transformations of the ``$\om$'' reactions since the transformations
of the ``$\op$'' reactions are their mirror images.

Our starting point, $\Psi_0$, is the reaction system with the defining
reactions
\begin{equation}
  \label{eq:14}
  \om \sigma w \rea[0] \phi(\sigma w) \om w,
  \qquad
  \sigma \in \Sigma, w \in \Sigma^{2r},
\end{equation}
(and their ``$\op$'' versions). Since $\Psi_0 = \Phi$ and $\hat\pi_b
\supseteq\bar\pi_b$, condition~\eqref{eq:wide-future} is true for
$\hat\pi$ too. All other properties of $\bar\pi$ in
Section~\ref{sec:unique-future} follow from~\eqref{eq:wide-future} and
are therefore also valid for $\hat\pi$. The most important of them is
that for all $a \in \Psi_0$ the process $\hat\pi_a$ exists and so
$\Psi_0$ is actually a reaction system. The following transformations
are defined in such a way that they preserve
property~\eqref{eq:wide-future}, which amounts to a proof that $\Psi$
is a valid reaction system.

\begin{enumerate}
\item To describe the first transformation we write $w \in
  \Sigma^{2r+1}$ as $\omega_0 \dots \omega_{2r}$ and define $w_{i,j} =
  \omega_i \dots \omega_j$. Then a typical defining reaction of
  $\Psi_0$ can be written as
  \begin{equation}
    \label{eq:psi0-typical}
    \om \omega_{0,2r} \rea[0] \sigma \om w_{1,2r}\,.
  \end{equation}
  In $\Psi_1$ it is replaced with reactions
  \begin{eqnarray}
    \label{eq:psi1-rea}
    \om_{r-k} \omega_{k,2r}
    &\rea[1]& \sigma \om_{r-\ell} w_{\ell+1,2r},\\
    \label{eq:psi1-superflouos}
    \om_{r-i} \omega_{i,2r-1}
    &\rea[1]& \om_{r-k} \omega_{k,2r-1}
    \qquad\quad\text{for all $i < k$}
  \end{eqnarray}
  for appropriate values of $k$ and $\ell$. We say then that the left
  side of~\eqref{eq:psi0-typical} has been \emph{reduced} by $k$ cells
  and the right side by $\ell$ cells.

  We can also say that these reductions have been achieved by applying
  the reaction $\om w_{0,2r-1} \rea[1] \om_{r-k} w_{k,2r-1}$ to the
  left side and $\om w_{1,2r} \rea[1] \om_{r-\ell} w_{k,2r}$ to the
  right side of~\eqref{eq:psi0-typical}. Both reactions are of
  type~\eqref{eq:psi1-superflouos}, and they will be noted in the
  calculations of Section~\ref{sec:narrow-form-110} to show what has
  been done.

  It remains to find values for $k$ and $\ell$. The transformation
  of~\eqref{eq:psi0-typical} is of course only sensible if the value
  of $w_{0,k-1}$ has no influence on $\sigma$. This means that there
  must be for all $\xi \in \Sigma$, $x \in \Sigma^{k-1}$ a reaction
  \begin{equation}
    \label{eq:all-superfluous}
    \om \xi x w_{k,2r} \rea[0] \sigma \om x w_{k,2r},
  \end{equation}
  and~\eqref{eq:psi1-superflouos} is a common replacement for all of
  them, or else $k = 0$. But it is also necessary that no cell is
  removed that is needed in following reactions. Therefore we must
  define $k$ and $\ell$ recursively:
  \begin{enumerate}
  \item $\ell$ is the largest value such that for every $\tau \in
    \Sigma$ the left side of the reaction in $\Psi_0$ that starts with
    $\om w_{1,2r} \tau$ can be reduced by at least $\ell$ cells, and
  \item $k$ is the largest value $\leq \ell - 1$ that
    fulfils~\eqref{eq:all-superfluous}.
  \end{enumerate}
  This induction can begin because a value of $\ell = 0$ is always
  possible.

  The reactions of~\eqref{eq:psi1-superflouos} are necessary because
  those of~\eqref{eq:psi1-rea} do not always match correctly: one of
  the reactions that start with $\om w_{1,2r} \tau$ may have been
  reduced by more than $\ell$ cells. In this case one of the reactions
  of~\eqref{eq:psi1-superflouos} removes the superfluous cells.

\item Next we \emph{unify} reactions that differ only on the right
  side.  If $k$ is maximal so that
  \begin{equation}
    \om_i uw \rea[1] \sigma \om_j vw
  \end{equation}
  for all $w \in \Sigma^k$, then
  \begin{equation}
    \om_i u \rea[2] \sigma \om_j v\,.
  \end{equation}
  Since $k$ can be $0$, every defining reaction in $\Psi_1$ has its
  counterpart in~$\Psi_2$.

\item It is possible that a defining reaction of $\Psi_2$ can be
  applied to the result of another defining reaction, and to its
  result possibly others. Then we have a sequence
  \begin{equation}
    \label{eq:21}
    \om_{j_0} u_0 \rea
    \sigma_1 \om_{j_1} u_1 \rea
    \dots \rea
    \sigma_1\dots \sigma_k \om_{j_k} u_k\,.
  \end{equation}
  Every defining reaction in $\Psi_2$ is the start of such a sequence,
  possibly of length 1. The length is always finite because the $u_i$
  never become longer and the maximal value of each $j_i$ is $r$.

  Therefore we can extend every reaction $\om_{j_0} u_0 \rea[2]
  \sigma_1 \om_{j_1} u_1$ to a maximal sequence~\eqref{eq:21} and set
  \begin{equation}
    \om_{j_0} u_0 \rea[3]
    \sigma_1\dots \sigma_k \om_{j_k} u_k\,.
  \end{equation}
  This gives the defining reactions of $\Psi_3$.

\item As a result of this and of~\eqref{eq:psi1-superflouos}, some
  defining reactions in $\Psi_3$ can never be applied to the result of
  another reaction. They have no influence on the long-term behaviour
  of the system and are therefore left out. The rest (and the
  corresponding ``$\op$'' reactions) are the defining reactions
  of~$\Psi$.
\end{enumerate}
This procedure has been defined in such a way that it always ends in a
finite number of steps, so questions of halting and computability do
not arise here.

\subsection{The Narrow Form of Rule 110}
\label{sec:narrow-form-110}

\begin{figure}[t]
  \def\l{\ar@{-}}
  \def\L{\ar@{=}}
  \def\r#1#2#3-#4{\sreaction{#1&#2&#3}{#4}}
  $$\xymatrix{
    & {\r\1\1\1-\0}
    \cr
      {\r\1\1\0-\1} \l[ur]
    & {\r\1\0\1-\1} \l[u]
    & {\r\0\1\1-\1} \l[ul]
    \cr
      {\r\1\0\0-\0} \l[u]  \l[ur]
    & {\r\0\1\0-\1} \L[ul] \L[ur]
    & {\r\0\0\1-\1} \L[ul] \L[u]
    \cr
    & {\r\0\0\0-\0} \L[ul] \l[u] \l[ur]
  }$$
  \caption{Rule 110 as a graph.  Every vertex describes a transition
    $\phi(w) = \sigma$.  Vertices that only differ by one cell in $w$
    are connected by a line.  They are connected by a double line if
    their values of $\sigma$ are equal.}
  \label{fig:lattice110}
\end{figure}
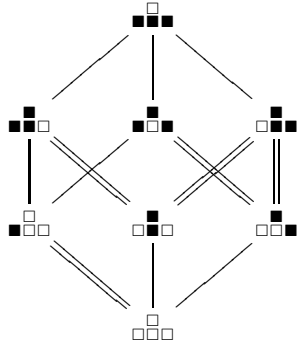
We can now read from Figure~\ref{fig:lattice110} the following cases
where the value of $\phi$ does not depend on all cells:
\begin{equation}
  \label{eq:15}
  \forall \sigma \in \Sigma\colon\,
  \phi(\sigma 00) = 0 \land
  \phi(\sigma 01) = 1 \land
  \phi(\sigma 10) = 1 \land
  \phi(01 \sigma) = 1
\end{equation}
These are the cases where the ignored cells are at the end, because
only they can be used here. They influence the constructions of
$\Psi_-$ and $\Psi_+$ in different ways.

In the case of $\Psi_-$, the first two terms in~\eqref{eq:15} lead to
the reactions $\om \sigma 00 \rea 0 \om 00$ and $\om \sigma 01 \rea 1
\om 01$.  It is therefore possible to reduce situations of the form
$\om \sigma 0$ to $\om_0 0$.  The last term in~\eqref{eq:15}, written
as a reaction, is $\om 01\sigma \rea 0\om 1 \sigma$: these two
reactions can be unified by committing the $\sigma$.

The computation that follows from these ideas can be summarised in the
following scheme:
\begin{displaymath}
  \def\pair#1#2{\left.\begin{array}{l}#1\\#2\\\end{array}\right\}}
  \def\[{\llap{$[$}}            
  \def\skip{\noalign{\medskip}}
  \begin{array}{l@{\qquad}llll@{\qquad}l}
    & \[\om 00 &\rea[a]& \om_0 0]\\
    &   \om 10 &\rea[b]& \om_0 0\\
    \pair{\om 000}{\om 100} \rea 0 \om 00 &
    \om_0 00 &\rea& 0 \om_0 0 &
    &\text{reduced by $\rea[a]$}\\\skip
    \pair{\om 001}{\om 101} \rea 0 \om 01 &
    \om_0 01 &\rea& 1 \om 01 &\rea[c] 11 \om 1
    &\text{reduced by $\rea[b]$}\\\skip
    \pair{\om 010 \rea 1 \om 10}{\om 011 \rea 1 \om 11} &
    \[\om 01 &\rea[c]& 1 \om 1]\\
    & \om 110 &\rea& 1 \om 10 & \rea[b] 1 \om_0 0\\
    & \om 111 &\rea& 1 \om 11
  \end{array}
\end{displaymath}
This diagram must be read from left to right. The first column
contains those generator reactions that can be transformed. The middle
column has, from top to bottom, the new reduction reactions, the
transformed reactions, and the untransformed reactions. If a reaction
can be continued, the result is appended at the right. The indices on
some of the reaction arrows are used only in this diagram and show
which of the reactions have been applied. Finally, the square brackets
mark reactions whose left side never occurs in a reaction result --
here because there is no reaction that creates $\om 0$.

The case of $\Psi_+$ is much simpler because there are only
unifications, name\-ly the three at the right of \eqref{eq:15}. Here
the computation is this:
\begin{displaymath}
  \def\pair#1#2{\left.\begin{array}{l}#1\\#2\\\end{array}\right\}}
  \def\skip{\noalign{\medskip}}
  \begin{array}{l@{\qquad}rlrl}
    \pair{000\op \rea 00\op 0}{100\op \rea 10\op 0}
    &00 \op &\rea   & 0 \op 0 \\\skip
    \pair{000\op \rea 00\op 0}{100\op \rea 10\op 0}
    &01 \op &\rlap{$\rea[a]$}& 0 \op 1 \\\skip
    \pair{010\op \rea 01\op 1}{110\op \rea 11\op 1}
    &10 \op &\rea& 1 \op 1 \\
    & 011 \op &\rea& 01 \op 1 & \rea[a] 0 \op 11 \\
    & 111 \op &\rea& 11 \op 1 \\
  \end{array}
\end{displaymath}
From these reactions all other reactions of $\Psi$ are derived. The
result is shown in Table~\ref{tab:narrow110}.
\begin{table}[t]
  \begin{center}
    \def\r{$&${}\rea}
    \begin{tabular}{l@{\quad}l}
      \toprule
      States: & 0, 1, $\om$, $\om_0$, $\op$.\\
      \midrule
      Situations: &
      \begin{tabular}[t]{@{}l@{ extends to }l}
        $\om$   & $\{\om 1\}$\\
        $\om_0$ & $\{\om_0 0\}$\\
        $\op$   & $\{0\op, 1\op\}$\\
      \end{tabular}\\
      \midrule
      Reactions: &
      \tabcolsep=0pt
      \begin{tabular}[t]{ll@{\qquad}rl}
        $\om_0 00 \r 0  \om_0 0$ & $00  \op \r 0  \op 0$ \\
        $\om_0 01 \r 11 \om 1$   & $10  \op \r 1  \op 1$ \\
        $\om 10   \r    \om_0 0$ & $01  \op \r 0  \op 1$ \\
        $\om 110  \r 1  \om_0 0$ & $011 \op \r 0  \op 11$ \\
        $\om 111  \r 0  \om 11$  & $111 \op \r 11 \op 0$ \\[0.5ex]

        $\hfill 0  \r 0 \op\om_0 0$ & $\om_0 0\op \r \lambda$ \\
        $\hfill 11 \r 11 \op\om 11$ & $\om 11\op \r \lambda$  \\
      \end{tabular} \\
      \bottomrule
    \end{tabular}
  \end{center}
  \caption{Rule 110, narrow form}
  \label{tab:narrow110}
\end{table}

\section{A Rule for Flexible Time}

Now we can return to the generalised time concept of the introduction.
I will choose a very simple kind of ``subroutines'' and show how to
take snapshots of the system when they have stopped.

These ``subroutines'' are all finite sequences of zeroes and have a
common behaviour,
\begin{equation}
  \label{psi-zeros}
  0^k \rea (0 \op)^k (\om_0 0)^k \qquad\text{for $k\geq 0$.} 
\end{equation}
They can be interpreted as a kind of timer which lasts $k$ time steps,
as many as the initial sequence is long. Geometrically,
\eqref{psi-zeros} traces a triangle of zeros, with its base at the
left side and the other edges at the right side of the reaction.

We can simplify the formulas by introducing the abbreviations
\begin{equation}
  \label{eq:epsilons}
  \e_- = \om_0 0, \qquad \e_+ = 0\op,
\end{equation}
then~\eqref{psi-zeros} becomes
\begin{equation}
  \label{eq:zeros}
  0^k \rea \epsilon_+^k \epsilon_-^k \qquad\text{for $k\geq 0$.} 
\end{equation}

In the new, third, reaction system we will have only situations where
all subprocesses~\eqref{eq:zeros} have ended. This means that the
elements of $\Sigma^+$ are not among the situations. But if we
apply~\eqref{eq:zeros} to all maximal subsequences of zeros then we
get from an element of $\Sigma^*$ a situation of the form
\begin{equation}
  \label{eq:epsilon-situation}
  1^{\ell_0}
  \epsilon_+^{k_1} \epsilon_-^{k_1} 1^{\ell_1}
  \dots
  \epsilon_+^{k_n} \epsilon_-^{k_n} 1^{\ell_n}
\end{equation}
with $\ell_0$, $\ell_n \geq 0$, and all other $k_i$, $\ell_i \geq 1$.
It has the additional property that it consists only of $1$,
$\epsilon_-$ and $\epsilon_+$, and that it does not contain
$\epsilon_- \epsilon_+$. We can now introduce reactions that preserve
this.

To get them, we evolve the situations $\epsilon_- 1^k \epsilon_+$ with
$k\geq 1$. Since the equivalent of~\eqref{eq:zeros} for ones is only
valid for sequences longer than one cell,
\begin{equation}
  \label{eq:ones}
  1^{k+2}
  \rea 11 \op 0^k \om 11
  \rea 11 \op \epsilon_-^k \epsilon_+^k \om 11
  \qquad\text{for $k\geq 0$,}
\end{equation}
we have to distinguish two cases:
\begin{equation}
  \begin{array}[b]{rll}
    \e_- 1 \e_+      &= \om_0 010 \op &\rea 11,\\
    \e_- 1^{k+2}\e_+ &= \om_0 0\, 1^{k+2}\, 0 \op
    &\rea \om_0 011\op \e_+^k \e_-^k \om 110\op\\
    &&\rea 11 \e_+^k \e_-^k 1\,.
  \end{array}
\end{equation}
The resulting reaction system is summarised in
Table~\ref{tab:selective110}. A way for further research into Rule~110
will be to search for other, more complex, subprocesses and
incorporate them into the reaction system as well.
\begin{table}[t]
  \begin{center}
    \begin{tabular}{l@{\quad}l}
      \toprule
      States: & $\epsilon_-$, $\epsilon_+$, 1.\\
      \midrule
      Situations:
      & No subsequence $\epsilon_- \epsilon_+$. \\
      \midrule
      Reactions: &
      \tabcolsep=0pt
      \begin{tabular}[t]{ll}
        $\e_- 1 \e_+$      & ${}\rea 1^2$, \\
        $\e_- 1^{k+2}\e_+$ & ${}\rea 1^2 \e_+^k \e_-^k 1$,
        \quad$k\geq 0$.
      \end{tabular} \\
      \bottomrule
    \end{tabular}
  \end{center}
  \caption{Rule 110, selective evolution}
  \label{tab:selective110}
\end{table}

\subsubsection{Acknowledgement} I want to thank Genaro Juárez
Martínez for encouragement and discussion.


\begin{thebibliography}{1}
\bibitem{Gosper1984} R. Wm. Gosper. Exploiting Regularities in Large
  Cellular Spaces. \emph{Physica \textbf{10}D (1984)}, p.~75--80.
\bibitem{MisnerThorneWheeler1973} Charles W. Misner, Kip S. Thorne,
  John Archibald Wheeler. \emph{Gravitation}. New York 1973. 
\bibitem{Wolfram1983} Stephen Wolfram. Statistical Mechanics of
  Cellular Automata. \emph{Reviews of Modern Physics} \textbf{55}
  (1983), p.~601--644.
\end{thebibliography}
\end{document}